\documentstyle[12pt]{article}
\begin{document}

\title{The Dirac equation in Cartesian gauge}

\author{Ion I. Cot\u aescu\\ {\it The West University of Timi\c soara,}\\
{\it V. Parvan Ave. 4, RO-1900 Timi\c soara}}

\maketitle

\begin{abstract}
It is shown that in the case of the spherically symmetric static backgrounds 
there is a gauge in which  the Dirac equation is manifestly covariant under  
rotations. This allows us to separate the spherical variables like in  the 
flat space-time, obtaining a pair of radial equations and a specific form of 
the radial scalar product.  

\end{abstract}
\

In the gauge field theory \cite{G} on curved space-time the physical meaning 
does not depend on the choice of the natural (holonomic) frame or on the gauge 
of the tetrad field which defines the local ones \cite{W,MTW}. However, from 
the technical point of view, these frames are not completely equivalent in the  
cases when the background has a global symmetry. Then, the symmetry 
transformations of the tetrad field are generally local, arising from an 
induced representation \cite{MAK} of the gauge group for which the group of the 
global symmetry play the role of a little group. Hence, in  such  gauges we 
have to face with the  difficulties of the theory of the induced 
representations which could hide some  properties, especially when the form of 
the field equations depends on the tetrad field, like in the Dirac theory. For 
this reason it seems that  the gauge in which the tetrad field  as well as the 
field equation  are manifestly covariant under the global symmetry (i.e. 
transforming according to linear representations) \cite{W} could offer certain 
technical advantages.

The spherical symmetric static backgrounds have the global symmetry of the 
group $T(1)\otimes SO(3)$, of the time translations and the rotations of the 
Cartesian space coordinates. There exist a gauge in which the tetrad field in 
spherical coordinates has only diagonal components  and another one where the 
tetrad field in Cartesian coordinates is manifestly covariant under rotations. 
This will be referred as the Cartesian gauge. Usually for deriving the Dirac 
equation one prefers the diagonal tetrad gauge where the result is obtained 
directly in spherical coordinates \cite{D,VIL}. Despite of the obvious 
advantages of  this gauge we believe that the study of the Dirac equation in 
the Cartesian gauge is also interesting. This will be done here.
          
We  show that in this gauge the Dirac equation can be put in a simple form by 
using an appropriate transformation of the spinor field \cite{VIL}.  Since this 
equation is manifestly covariant under rotations we can use  the known results 
from the special relativity  in order to separate the spherical 
variables in terms of the angular momentum eigenspinors \cite{BJDR,TH}. 
Thus we obtain the radial equations and the form of the radial scalar product 
in the most general case of a spherically symmetric static metric. We shall 
work in natural units, $\hbar=c=1$.    

Let us consider a background where we have introduced the natural (holonomic) 
frame of the coordinates $x^{\mu}, \mu=0,1,2,3$.  We denote by $e_{\hat\mu}(x)$ 
the tetrad fields which define the local frames and by $\hat e^{\hat\mu}(x)$ 
that of the corresponding coframes. These have the usual orthonormalization 
properties   
\begin{equation}
e_{\hat\mu}\cdot e_{\hat\nu}=\eta_{\hat\mu \hat\nu}, \quad
\hat e^{\hat\mu}\cdot \hat e^{\hat\nu}=\eta^{\hat\mu \hat\nu}, \quad 
\hat e^{\hat\mu}\cdot e_{\hat\nu}=\delta^{\hat\mu}_{\hat\nu}, 
\end{equation}
where $\eta=$diag$(1,-1,-1,-1)$ is the Minkowki metric. The 1-forms of the local 
frames,  $d\hat x^{\hat\mu}=\hat e_{\nu}^{\hat\mu}dx^{\nu}$, allow one to write 
the line element  
\begin{equation}\label{(met)}
ds^{2}=\eta_{\hat\mu \hat\nu}d\hat x^{\hat\mu}d\hat x^{\hat\nu}=
g_{\mu \nu}(x)dx^{\mu}dx^{\nu}.
\end{equation}   
which defines the metric tensor $g_{\mu \nu}$ of the natural frame. This raises 
or lowers the Greek indices (ranging from 0 to 3) while for the hat Greek ones 
(with the same range) we have to use the Minkowski metric, 
$\eta_{\hat\mu \hat\nu}$. The derivatives  
$\hat\partial_{\hat\nu}=e^{\mu}_{\hat\nu}\partial_{\mu}$ satisfy the 
commutation rules  
\begin{equation}
[\hat\partial_{\hat\mu},\hat\partial_{\hat\nu}]
=e_{\hat\mu}^{\alpha} e_{\hat\nu}^{\beta}(\hat e^{\hat\sigma}_{\alpha,\beta}-
\hat e^{\hat\sigma}_{\beta,\alpha})\hat\partial_{\hat\sigma}
=C_{\hat\mu \hat\nu 
\cdot}^{~\cdot \cdot \hat\sigma}\hat\partial_{\hat\sigma}
\end{equation}
defining the Cartan coefficients which halp us to write the conecttion 
components in the local frames as   
\begin{equation}
\hat\Gamma^{\hat\sigma}_{\hat\mu \hat\nu}=e_{\hat\mu}^{\alpha}
e_{\hat\nu}^{\beta}
(\hat e^{\hat\sigma}_{\beta, \alpha}+\hat e_{\gamma}^{\hat\sigma}
\Gamma^{\gamma}_{\alpha \beta})=
\frac{1}{2}\eta^{\hat\sigma \hat\lambda}(C_{\hat\mu \hat\nu \hat\lambda}+
C_{\hat\lambda \hat\mu \hat\nu}+C_{\hat\lambda \hat\nu \hat\mu})
\end{equation}
while the notation  $\Gamma^{\gamma}_{\alpha \beta}$ stands for the usual 
Christoffel symbols.

Let $\psi$ be a Dirac free field of the mass $M$, defined on the   
space domain $D$.  This has the gauge invariant action \cite{BD}     
\begin{equation}\label{(action)}
{\cal S}[\psi]=\int_{D} d^{4}x\sqrt{-g}\left\{
\frac{i}{2}[\bar{\psi}\gamma^{\hat\alpha}D_{\hat\alpha}\psi-
(\overline{D_{\hat\alpha}\psi})\gamma^{\hat\alpha}\psi] - 
M\bar{\psi}\psi\right\}
\end{equation}
where 
\begin{equation}
D_{\hat\alpha}=\hat\partial_{\hat\alpha}+\frac{i}{2}S^{\hat\beta \cdot}_{\cdot 
\hat\gamma}\hat\Gamma^{\hat\gamma}_{\hat\alpha \hat\beta},
\end{equation}
are the covariant derivatives of the spinor field  
and $g=\det(g_{\mu\nu})$. 
The Dirac matrices,  $\gamma^{\hat\alpha}$, and the generators of the reducible 
spinor representation of the $SL(2,C)$ group, $S^{\hat\alpha \hat\beta}$, 
satisfy
\begin{eqnarray}
&&\{ \gamma^{\hat\alpha}, \gamma^{\hat\beta} \}=2\eta^{\hat\alpha \hat\beta}, 
\qquad [\gamma^{\hat\alpha}, \gamma^{\hat\beta} ]=-4iS^{\hat\alpha \hat\beta},\\ 
&&~~~~~~~[ S^{\hat\alpha \hat\beta}, \gamma^{\hat\mu}]=
i(\eta^{\hat\beta \hat\mu}\gamma^{\hat\alpha}-
\eta^{\hat\alpha \hat\mu}\gamma^{\hat\beta}).
\end{eqnarray}
Thereby it results that the field equation,
\begin{equation}\label{(d)}
i\gamma^{\hat\alpha}D_{\hat\alpha}\psi - M\psi=0,
\end{equation}
derived from  (\ref{(action)})  can be written as 
\begin{equation}\label{(dd)}
i\gamma^{\hat\alpha}e_{\hat\alpha}^{\mu}\partial_{\mu}\psi - M\psi
+ \frac{i}{2} \frac{1}{\sqrt{-g}}\partial_{\mu}(\sqrt{-g}e_{\hat\alpha}^{\mu})
\gamma^{\hat\alpha}\psi
-\frac{1}{4}
\{\gamma^{\hat\alpha}, S^{\hat\beta \cdot}_{\cdot \hat\gamma} \}
\hat\Gamma^{\hat\gamma}_{\hat\alpha \hat\beta}\psi =0.
\end{equation}
On the other hand, from the conservation of the electric charge we can deduce 
that when $e^{0}_{i}=0$,  $i=1,2,3$, then the time-independent relativistic 
scalar product of two spinors is
\cite{BD}
\begin{equation}\label{(sp)}
(\psi,\psi')=\int_{D}d^{3}x\mu(x)\bar\psi(x)\gamma^{0}\psi'(x), \quad 
\mu=\sqrt{-g}e_{0}^{0}.
\end{equation}

Our aim  is to discuss here only the case of the  spherically symmetric static 
backgrounds which, as mentioned, have the global symmetry of the $T(1)\otimes 
SO(3)$ group. These have natural frames of the Cartesian coordinates   
$x^{0}=t$ and  $x^i$, $i=1,2,3$, in which  the metric tensor is 
time-independent and manifestly covariant  under the  rotations $R\in SO(3)$ of 
the space coordinates,
\begin{equation}\label{(rot)}  
x^{\mu}\to x'^{\mu}=(Rx)^{\nu} \qquad
(t'=t,\quad  x'^{i}= R_{ij}x^{j}).
\end{equation}
The most general form of a such a metric is given by the line element     
\begin{equation}\label{(metr)} 
ds^{2}=g_{\mu\nu}(x)dx^{\mu}dx^{\nu}=A(r)dt^{2}-[B(r)\delta_{ij}+C(r)x^{i}x^{j}]
dx^{i}dx^{j}
\end{equation} 
where  $A$, $B$ and $C$  are arbitrary functions of the Euclidian norm of 
$\stackrel{\rightarrow}{x}$,  $r=\vert\stackrel{\rightarrow}{x}\vert$ (which is 
invariant under rotations).  In applications it is convenient to replace these 
functions by  new ones,  $u$,  $v$ and $w$, such that
\begin{equation}\label{(ABC)}   
A=w^{2}, \quad B=\frac{w^2}{v^2}, \quad 
C=\frac{w^2}{r^2}\left( \frac{1}{u^2}-\frac{1}{v^2}\right).
\end{equation}
Then the metric appears as the conformal transformation of that simpler one 
having $w=1$.

Starting with a Cartesian natural frame we define the Cartesian gauge in which                                                            
the static tetrad field  transforms under the rotations (\ref{(rot)}) 
according to the rule
\begin{equation}\label{(tr)}
d\hat x ^{\hat\mu}\to d\hat x'^{\hat\mu}=\hat e^{\hat\mu}_{\alpha}(x')dx'^{\alpha}
=(Rd\hat x)^{\hat\nu}.
\end{equation}
In the case of the metric (\ref{(metr)}) the simplest choice of their 
components is
\begin{eqnarray}
\hat e^{0}_{0}&=&\hat a(r), \quad \hat e^{0}_{i}=\hat e^{i}_{0}=0, \quad
\hat e^{i}_{j}=\hat b(r)\delta_{ij}+\hat c(r) x^{i}x^{j},\label{(eee)}\\
e^{0}_{0}&=& a(r), \quad  e^{0}_{i}= e^{i}_{0}=0, \quad
e^{i}_{j}= b(r)\delta_{ij}+ c(r) x^{i}x^{j},\label{(eee1)}
\end{eqnarray}
where, according to (\ref{(met)}), (\ref{(metr)}) and (\ref{(ABC)}), we must 
have 
\begin{eqnarray}
\hat a&=&w, \quad \hat b=\frac{w}{v}, \quad \hat c=\frac{1}{r^2}
\left( \frac{w}{u}-\frac{w}{v}\right), \label{(abc)}\\
a&=& \frac{1}{w}, \quad  b=\frac{v}{w}, \quad  c=\frac{1}{r^2}
\left( \frac{u}{w}-\frac{v}{w}\right),\label{(abc1)}
\end{eqnarray}                                                                                                                                                                                                                                                 
      
while the weight function of (\ref{(sp)}) becomes
\begin{equation}\label{(mu)}
\mu
=\frac{1}{b^{2}(b+r^{2}c)}
=\frac{w^3}{uv^2}
\end{equation} 
since
\begin{equation}
\sqrt{-g}=B[A(B+r^{2}C)]^{1/2}=\frac{1}{ab^{2}(b+r^{2}c)}
=\frac{w^4}{uv^2}.
\end{equation}

Now we have to replace the concrete form of the tetrad components in 
Eq.(\ref{(dd)}). First we eliminate its last term 
since it is known that this can not contribute when the metric is spherically 
symmetric. The argument is that 
$\{\gamma^{\hat\alpha},S^{\hat\beta \hat\gamma}\}=
\varepsilon^{\hat\alpha \hat\beta \hat\gamma \cdot}_{~~~ \hat\lambda}
\gamma^{5}\gamma^{\hat\lambda}$ (with $\varepsilon^{0123}=1$) is 
completely antisymmetric while 
the Cartan coefficients resulted from (\ref{(eee)}) and  (\ref{(eee1)}) have 
no such components. Furthermore,  in order to simplify the remaining equation, 
we perform the transformation
\begin{equation}\label{(cfu)}
\psi(x)=\chi(r)\hat\psi(x).
\end{equation}
If we chose 
\begin{equation}\label{(chi)}
\chi=[\sqrt{-g}(b+r^{2}c)]^{-1/2}=b\sqrt{a}=vw^{-3/2},     
\end{equation}
then all the terms containing the derivatives of the functions $a$ and  $b$   
are eliminated (while that of $c$ is not involved). Thus  we obtain the Dirac 
equation in Cartesian gauge,
\begin{equation}\label{(de)}
i\{a(r)\gamma^{0}\partial_{t} +b(r)(\stackrel{\rightarrow}{\gamma}\cdot 
\stackrel{\rightarrow}{\partial})+
c(r)(\stackrel{\rightarrow}{\gamma}\cdot \stackrel{\rightarrow}{x})[1+
(\stackrel{\rightarrow}{x}\cdot \stackrel{\rightarrow}{\partial})]\}\hat\psi(x)
-M\hat\psi(x)=0,
\end{equation}
expressed only on familiar three-dimensional scalar products and these 
functions of $r$. It is clear that this equation is manifestly covariant under 
rotation and, consequently, all the properties related to the conservation of 
the angular momentum, including the separation of the variables in spherical 
coordinates, will be similar as those of the usual  Dirac theory in the 
Minkowski flat space-time.

By using (\ref{(abc1)}) and the traditional notations, 
$\alpha^{i}=\gamma^{0}\gamma^{i}$ and 
$\beta=\gamma^{0}$, we  can  write the Hamiltonian form of this equation as  
\begin{equation}\label{(eham)}
(\hat H \hat\psi )(x)=i\partial_{t}\hat\psi (x),
\end{equation}
where the operator $\hat H$ with the action
\begin{eqnarray}\label{(ham)}
&&(\hat H \hat\psi )(x)=\label{(ham)}\\
&&-i\left\{ v(r) (\stackrel{\rightarrow}{\alpha}\cdot 
\stackrel{\rightarrow}{\partial})+
\frac{u(r)-v(r)}{r^2}(\stackrel{\rightarrow}{\alpha}\cdot 
\stackrel{\rightarrow}{x})[1+
(\stackrel{\rightarrow}{x}\cdot \stackrel{\rightarrow}{\partial})]\right\}
\hat\psi(x)+w(r)M\beta\hat\psi(x)\nonumber
\end{eqnarray}
is the Hamiltonian  of the transformed Dirac field $\hat\psi$.
Obviously, according to (\ref{(cfu)}), the Hamiltonian  of the field 
$\psi$ is $H=\chi \hat H \chi^{-1}$. In our opinion,   
new interesting particular cases could be anlysed by using this form of the 
Dirac equation. However, here we restrict 
ourselves to note only that from (\ref{(ham)}) we can recover the known result 
that in the massless limit the Eq.(\ref{(eham)}) becomes invariant under the 
conformal transformations $g_{\mu \nu}\to \hat w^{2}g_{\mu\nu}$ and 
$\hat\psi\to \hat w^{-3/2}\hat\psi$ \cite{BD} where $\hat w$ is an  arbitrary 
function of $r$.

The last step is to introduce the spherical coordinates, $r$, $\theta$, $\phi$,  
associated with  the space  coordinates of our natural Cartesian frame.  
Doing this we obtain the line element   
\begin{equation}\label{(muvw)}
ds^{2}=w^{2}dt^{2}-\frac{w^2}{u^2}dr^{2}-
\frac{w^2}{v^2}r^{2}(d\theta^{2}+\sin^{2}\theta d\phi^{2}).
\end{equation}
The form of the Hamiltonian (\ref{(ham)}) allows us to 
separate the variables in Eq.(\ref{(eham)}) as 
in the case of the central motion in flat space-time, by using the 
four-components angular spinors $\Phi^{\pm}_{m_{j}, \kappa_{j}}(\theta, \phi)$ 
as given in Ref.\cite{TH}. These are orthogonal to each 
other being completely determined by the quantum number $j$ of the  angular 
momentum, the quantum number $m_{j}$ of its projection along the third axis 
and the value of $\kappa_{j}=\pm (j+1/2)$.   
Looking for  particular (particle-like) solutions of the form
\begin{equation}\label{(psol)}
\hat\psi_{E,j,m_{j},\kappa_{j}}(t,r,\theta,\phi)
=\frac{1}{r}[f_{+}(r)\Phi^{+}_{m_{j},\kappa_{j}}(\theta,\phi)
+f_{-}(r)\Phi^{-}_{m_{j},\kappa_{j}}(\theta,\phi)]e^{-iEt},
\end{equation}
after a little calculation, we find  the desired radial equations
\begin{eqnarray}
\left[u(r)\frac{d}{dr}+v(r)\frac{\kappa_j}{r}\right]f_{+}(r)&=&[E+w(r)M
]f_{-}(r),\\
\left[- u(r)\frac{d}{dr}+v(r)\frac{\kappa_j}{r}\right]f_{-}(r)&=&[E-w(r)M
]f_{+}(r).
\end{eqnarray}
Thus, we have shown that in the Cartesian gauge we can separate  the radial 
motion from the angular one which can be completely solved grace to the 
conservation of the angular momentum. Like in the special relativity, the 
radial motion here is described by a pair of radial equations which must be 
solved in each particular case separately. In practice these can be written 
directly by starting with the metric (\ref{(muvw)}) from  which we can  
identify  the functions $u$, $v$, and $w$ we need.

The angular spinors are normalized so that the angular integral of the scalar 
product (\ref{(sp)}) does not contribute and, consequently, this reduces to a 
radial integral. By using (\ref{(cfu)}) and (\ref{(psol)}) we find that this is
\begin{equation}\label{(spp)}
(\psi,\psi')=\int_{D_{r}}\frac{dr}{u(r)}[{f_{+}}^{*}(r)f'_{+}
(r)+{f_{-}}^{*}(r)f'_{-}(r)]
\end{equation}
where $D_{r}$ is the radial domain corresponding to $D$. What  is remarkable 
here is that the weight function $\mu \chi^{2}=1/u$, resulted from (\ref{(mu)}) 
and (\ref{(chi)}), is just that we need in order to have 
$(u\partial_{r})^{+}=-u\partial_{r}$. 
Therefore, the operators of the left-hand side of our radial equations are 
related between them through the Hermitian conjugation with respect to 
(\ref{(spp)}). The direct consequence of this property is that 
the corresponding radial Hamiltonian operator is self-adjoint. Moreover, we 
observe that if there 
exists two real constants, $c_{1}$ and $c_{2}$, so that 
\begin{equation}\label{(vw)}
v=r(c_{1}+c_{2}w),
\end{equation}
then this is  supersymmetric. Indeed, one can easily 
verify that  a simple rotation in the plane $(f_{+},f_{-})$ is enough  to 
bring it in the canonical form  of a Hamiltonian 
with supersymmetry \cite{TH}. On the other hand, the metrics which satisfy 
(\ref{(vw)}) differ from the general ones  given by (\ref{(muvw)}) only by a 
conformal transformation. Thus we can conclude that the central motion 
of the Dirac particle in the 
general relativity is up to a conformal transformation  a problem with 
supersymmetry.


\begin{thebibliography}{20}

\bibitem{G} 
R. Utiyama, Phys. Rev. {\bf 101}, 1597 (1956);
T. W. B. Kibble, J. Math. Phys. {\bf 2}, 212 (1961)
 
\bibitem{W}
S. Weinberg, {\it Gravitation and Cosmology: Principles and Applications of 
the General Theory of Relativity}, Wiley, New York, 1972


\bibitem{MTW}
C. M. Misner, K. S. Thorne and J. A. Wheeler, {\it Gravitation}, W. H. Freeman 
\& Co., San Francisco, 1973 

\bibitem{MAK}
G. Mackey, {\it Induced Representations of Groups and Quantum Mechanics}, 
Benjamin, New York, 1968


\bibitem{D} 
D. R. Brill and J. A. Wheeler, Rev. Mod. Phys. {\bf 29}, 465 (1957);  
D. R. Brill and J. A. Cohen, J. Math. Phys. {\bf 7}, 238 (1966); 
J. Klauder and J. A. Wheeler, Rev. Mod. Phys. {\bf 29}, 516 (1957); 
T. M. Davis and J. R. Ray, J. Math. Phys. {\bf 16}, 75 (1975),
Phys. Rev.   {\bf D9}, 331 (1974),
J. Math. Phys. {\bf16}, 80 (1975); 
K. D. Kriori and H. Kakati, GRG {\bf 20}, 1237 (1995);
J. C. Huang, N. O. Santos and Kleber, Class. Quantum Grav. 
{\bf 12}, 1245 (1995);
I. D. Soares and J. Tiomno, Phys. Rev.  {\bf D54}, 2808 (1996);
C. G. De Oliveira and J. Tiomno, Il Nouvo Cimento {\bf 24}, 672 (1962);
B. D. B. Figueredo, I. D. Soares and Tiomno, Class. Quantum Grav. 
{\bf 9}, 1593 (1992);
Hammond R., Class. Quantum Grav. {\bf 12}, 279 (1995);
P. Baekler, M. Setz, V. Winkelmann, Class. Quantum Grav. {\bf 5}, 479 (1988);
I. I. Cot\u aescu and D. Vulcanov, Int. J. Mod. Phys. C {\bf 8}, 273 (1997)

\bibitem{VIL}
V. M. Villalba, preprint gr-qc/9306019 

\bibitem{BJDR} 
J. D. Bjorken and S. D. Drell S.D.  {\it Relativistic Quantum Mechanics}, 
McGraw-Hill Book Co., NY, 1964

\bibitem{TH} 
B. Thaller,  {\it The Dirac Equation}, Springer Verlag, Berlin 
Heidelberg, 1992



\bibitem{BD}
N. D. Birrel and P. C. W. Davies, {\it Quantum Fields in Curved Space}, 
Cambridge University Press, Cambridge (1982)

 
\end{thebibliography}
\end{document}